\documentclass{moriond}

\def\be{\begin{equation}}
\def\ee{\end{equation}}
\def\bea{\begin{eqnarray}}
\def\eea{\end{eqnarray}}

\usepackage[dvipsnames]{xcolor}
\usepackage{amsmath}
\newcommand{\X}{{\mathcal{X}}}

\newcommand{\Photo}{\includegraphics[height=35mm]{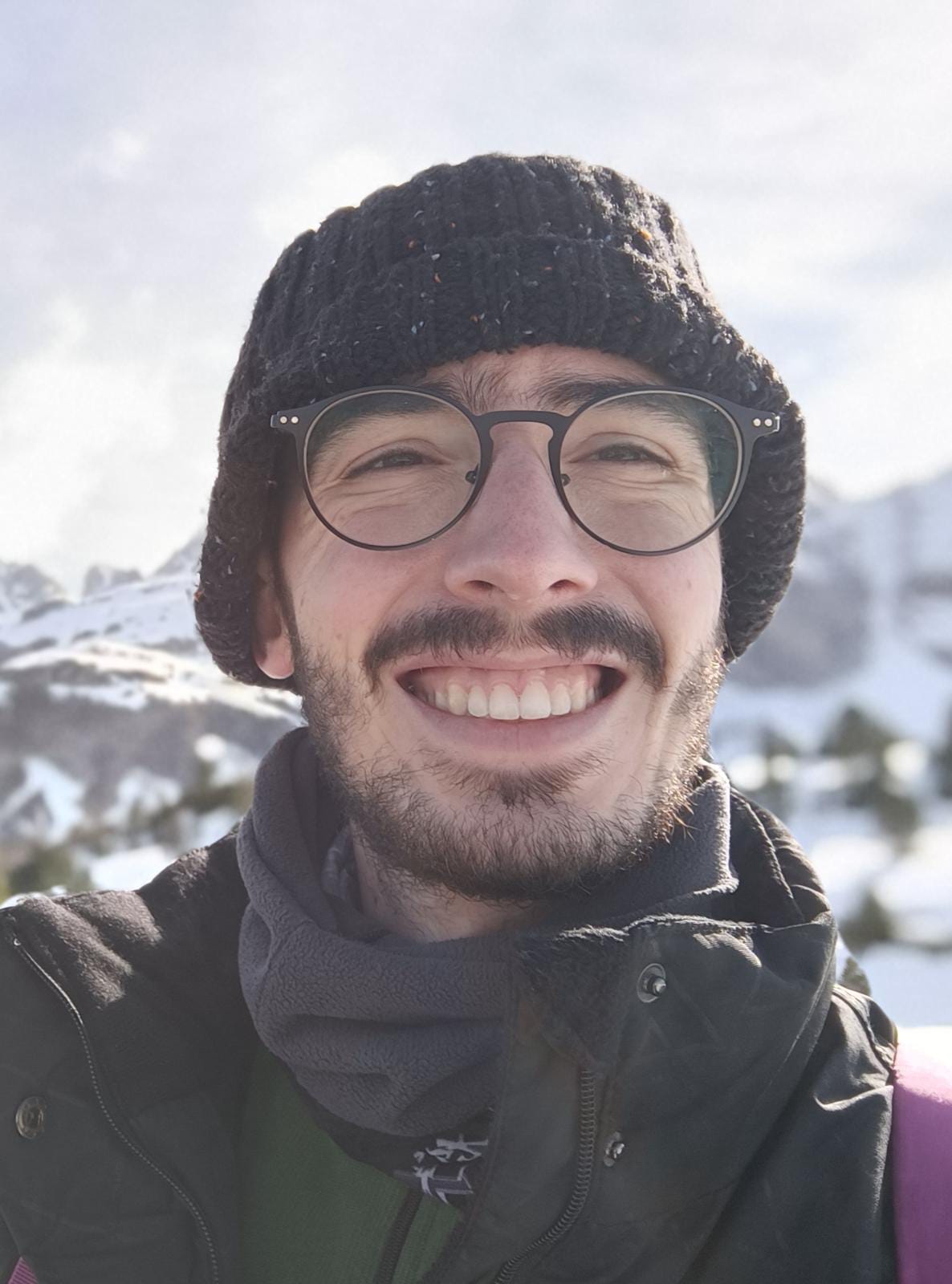}}

\begin{document}
\vspace*{4cm}
\title{UV-IR interplay in axion flavour violation}

\author{ Luca Di Luzio, Alfredo Walter Mario Guerrera, \\ Xavier Ponce D\'iaz \footnote{Speaker}, Stefano Rigolin}

\address{Istituto Nazionale di Fisica Nucleare (INFN), Sezione di Padova, \\Dipartimento di Fisica e Astronomia `G.~Galilei', Universit\`a di Padova,
 \\ Via F. Marzolo 8, 35131 Padova, Italy  }

\maketitle\abstracts{
Flavour-violating axions appear in models where the Peccei-Quinn(PQ) charges are generation non-universal. Consequently, this charge arrangement will also generate flavour violation in the UV sector. The typical way of implementing such an axion in a UV completion is with a DFSZ model, containing 2 Higgs doublets. In this talk we will present how to parameterize the flavour violation in the UV such that we can find a direct correlation with the flavour violation of the axion. Finally, we show in an example how this connection can help in giving information about the UV completion if an axion is found in a flavour-violating channel.}

\section{Introduction}

A flavour-violating axion (IR-source) appears when the $U(1)_{\textrm{PQ}}$ couples non-universally (i.e.~in a generation-dependent way) to the Standard Model (SM) fermions. Such type of axion models can be motivated in several respects: i) the connection to the Flavour puzzle \cite{Ema:2016ops,Calibbi:2016hwq,Arias-Aragon:2017eww}; ii) the possibility of achieving an ``astrophobic'' scenario \cite{DiLuzio:2017ogq,Bjorkeroth:2019jtx,DiLuzio:2022tyc} (where nucleon and electron couplings are suppressed); and iii) the opportunity of finding the axion in flavour-violating searches \cite{MartinCamalich:2020dfe,Calibbi:2020jvd,Jho:2022snj}. 
These previous points have in common that a flavour-violating axion must be accompanied by a UV completion, which in general is also flavour-violating. Here, we show how in a class of models, such as a non-universal DFSZ models, we can correlate in a one-to-one way the flavour violating observables of the axion with those of the UV completion, in this case a PQ-2 Higgs Doublet Model (PQ-2HDM). We use this correspondence between IR and UV observables in a practical example of what information could be inferred from the UV if an axion is observed in a flavour-violating golden channel, e.g.~$K\to\pi a$.

\section{Flavour violation in axion models}\label{subsec:FVgeneral}

\subsection{Flavour violation in the IR: the axion}\label{subsec:FVaxion}

Axions coupling to the SM fermions arise as complex phases in the mass matrix. Typically this term is removed by a field redefinition along the charges of the SM fermions, appearing as a derivative coupling of the form
\begin{equation}
    \mathcal{L}_{a} = \sum_f\frac{\partial_\mu a}{f_a}\, \bar f\, \X_f\gamma^\mu f\, , \quad \textrm{with } f=\{q_L,\,d_R,\,u_R,\, \ell_L,\,e_R\}\,.
\end{equation}
Here, the couplings $\X_f$ are the charges of the SM fermions with respect to the $U(1)_{\rm PQ}$ in flavour space. After rotation of the fermions into the mass basis, one finds that to have flavour violation in the axion sector it is necessary to have non-universal charges, or conversely, the charges must not be proportional to the identity,
\begin{equation}
    \frac{\partial_\mu a}{f_a}\, \bar f\, \X_f\gamma^\mu f \quad \longrightarrow \quad \frac{\partial_\mu a}{f_a}\, \bar f\, V_{f}^\dagger \X_f V_{f}\gamma^\mu f\, .
\end{equation}
The matrices $V_f$ correspond to the bi-unitary transformations which diagonalize the fermion mass matrix $\hat{M}_f=V_{f_L}^\dag M_f V_{f_R}$. It is useful to define these couplings as
\begin{equation}
    C^{R,L}_f = \frac{1}{2N} V_{f_{R,L}}^\dag \X_{f_{R,L}} V_{f_{R,L}}\, ,
\end{equation}
where $N$ is the QCD anomaly coefficient.
\subsection{Flavour violation in the UV: the PQ-2HDM}\label{subsec:the2HDM}

Generally, this axion can be generated as a variation of the classic DFSZ model \cite{Zhitnitsky:1980tq,Dine:1981rt}, implemented with non-universal charges. This model contains 2 Higgs doublets, which introduce an additional Yukawa matrix coupling to the fermions
\begin{equation}
    \mathcal{L}_{\rm PQ-2HDM} = \bar f_L \left(Y_1^f \tilde{H}_1+Y_2^f \tilde{H}_2\right) f_R +\rm h.c. \, .
\end{equation}

In this type of models one cannot generally diagonalize simultaneously the mass matrix and individual Yukawas, leading to the appearance of flavour violation in the UV-completion. This can be seen by noting that the mass matrix is a combination of the Yukawas and vacuum expectation values of the two Higgs, $v_1=v c_\beta$ and $v_2=v s_\beta$ with $s_\beta\equiv \sin\beta$, $c_\beta\equiv \cos\beta$ and $t_\beta \equiv \tan\beta=\frac{v_2}{v_1}$,
\begin{equation}
\label{eq:FermionMassMatrix}
    M_f=\frac{v}{\sqrt{2}}\left(Y_1^f\, c_\beta + Y_2^f\, s_\beta\right)\, ,
\end{equation}
while the interactions of the radial modes follow different combinations (see reference~\cite{DiLuzio:2023ndz}), generating flavour-violating couplings. However, there is no apparent direct connection between the flavour violation in the IR and UV sectors. Such a connection can be made explicitly manifest by noting that the structure of the Yukawas is given by the charge arrangement. Thus, by imposing the $U(1)_{\rm PQ}$ on the Yukawa matrices, one can obtain the following sum rules,
\begin{equation}
    -\X_{f_L} Y_{1,2}^f + Y_{1,2}^f \,\X_{f_R} + \X_{1,2} Y_{1,2}^f = 0 \, .
\end{equation}

These expressions combined with the definition of the mass matrix in Eq.~\ref{eq:FermionMassMatrix}, fix the specific shape of the Yukawas 
\begin{equation}
\label{eq:FVconnection}
    Y_1^d = \frac{\sqrt{2}}{v c_\beta}V_{d_L} \left( C^L_d \hat{M}_d - \hat{M}_d C^R_d + \X_2 \hat{M}_d \right)V^\dagger_{d_R} \, , \,Y_2^d = \frac{\sqrt{2}}{v s_\beta} V_{d_L}\left( - C^L_d \hat{M}_d + \hat{M}_d C^R_d - \X_1 \hat{M}_d \right)V^\dagger_{d_R} \, , 
\end{equation}
 where we took the specific case of the $d-$type quarks. Similar expressions hold for the $u-$type quarks and leptons \footnote{These expressions are given for $2N=1$ axion models.}. These relations, are the main result of reference \cite{DiLuzio:2023ndz}, and it is important to emphasize several points: i) the structure of flavour violation is contained in the couplings $C_f^{R,L}\sim V_{R,L}^\dagger \X_{f_{R,L}}V_{R,L}$, and it is shared across scales; ii) these relations don't assume any particular PQ-charge or Yukawa structure (unlike in other works where Yukawa \textit{ansatzs} are used); and iii) the charges of the Higgs will not play any role in flavour-violating couplings. 

\section{An example of the flavour connection across scales}

To emphasize the importance of these relations, we give a particular example of the information one could obtain of the UV-completion after a positive detection of an axion in the golden channels: $K\to \pi \, a$ and $B \to K \, a$. 

Using Eq.~\ref{eq:FVconnection} it is possible to relate observables of flavour-violating axions and the PQ-2HDM. To simplify, we use models with $2+1$ charge arrangements \cite{DiLuzio:2017ogq}, which exhibit flavour violation in just one of the two couplings, $C^{R,L}_f$. This allows to write the following direct relations which link the axion golden channels with the associated 2HDM meson mixing contribution in the alignment limit \cite{DiLuzio:2023ndz}
\begin{align}
 \label{eq:rel1}
 \left(\frac{f_a}{10^{11}\, \textrm{GeV}}\right)^2\left(\frac{1\, \textrm{TeV}}{s_{2\beta}\, m_H}\right)^2\left(\frac{\textrm{BR}(K\to\pi a)}{7.3\cdot10^{-11}}\right)=\frac{2\,|M^{\textrm{NP}}_{12}|}{3.5\cdot10^{-15}\, \textrm{GeV}} \, , \\
  \label{eq:rel2}
 \left(\frac{f_a}{8.8\cdot 10^7\, \textrm{GeV}}\right)^2\left(\frac{1\, \textrm{TeV}}{s_{2\beta}\, m_H}\right)^2\left(\frac{\textrm{BR}(B\to K a)}{7.1\cdot10^{-6}}\right)=\frac{2\, |M^{\textrm{NP}}_{12}|}{1.2\cdot10^{-11}\, \textrm{GeV}} \, .
\end{align}

\begin{figure}[t]
\begin{minipage}{0.49\linewidth}
\centerline{\includegraphics[width=\linewidth]{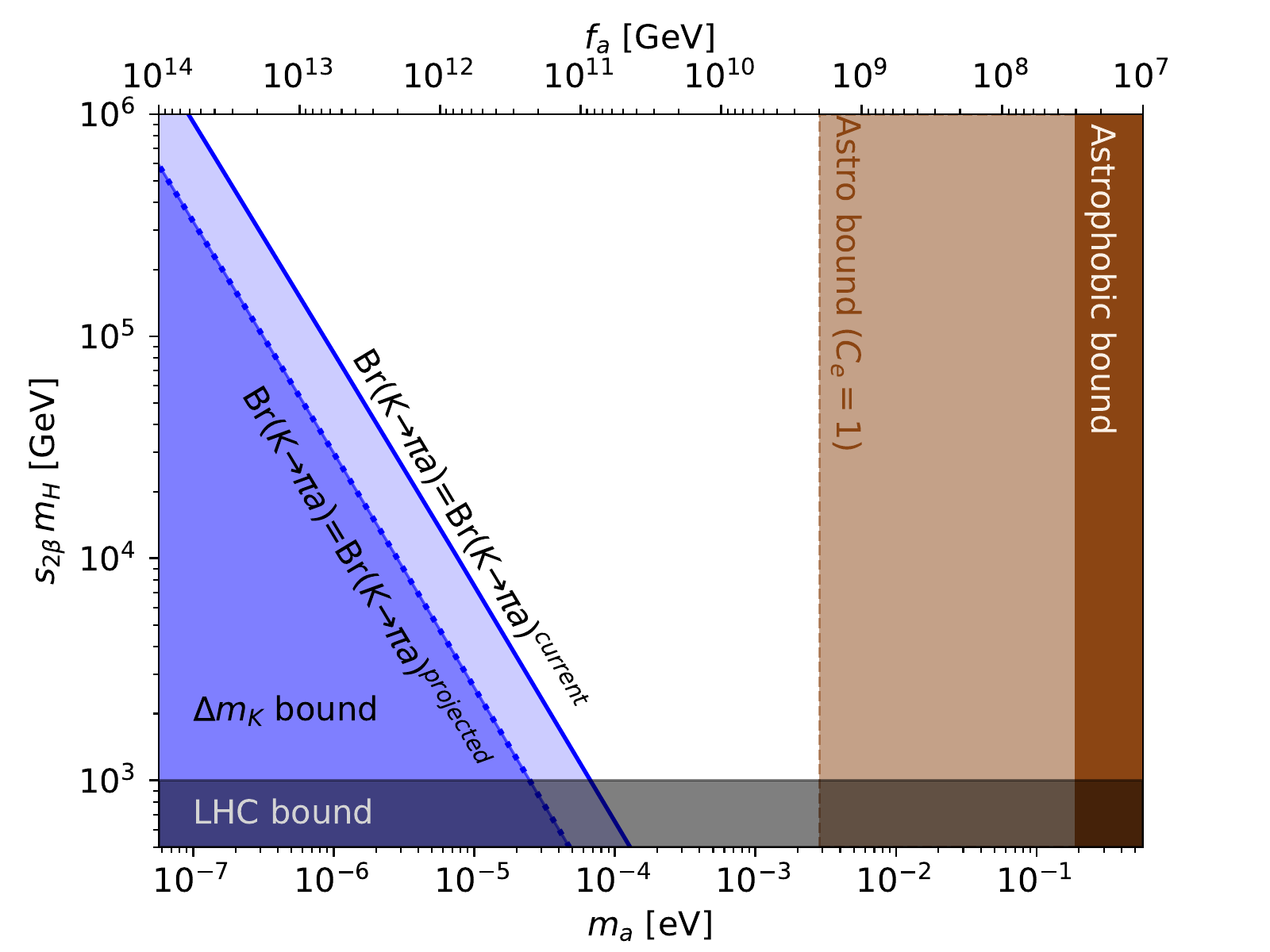}}
\end{minipage}
\hfill
\begin{minipage}{0.49\linewidth}
\centerline{\includegraphics[width=\linewidth]{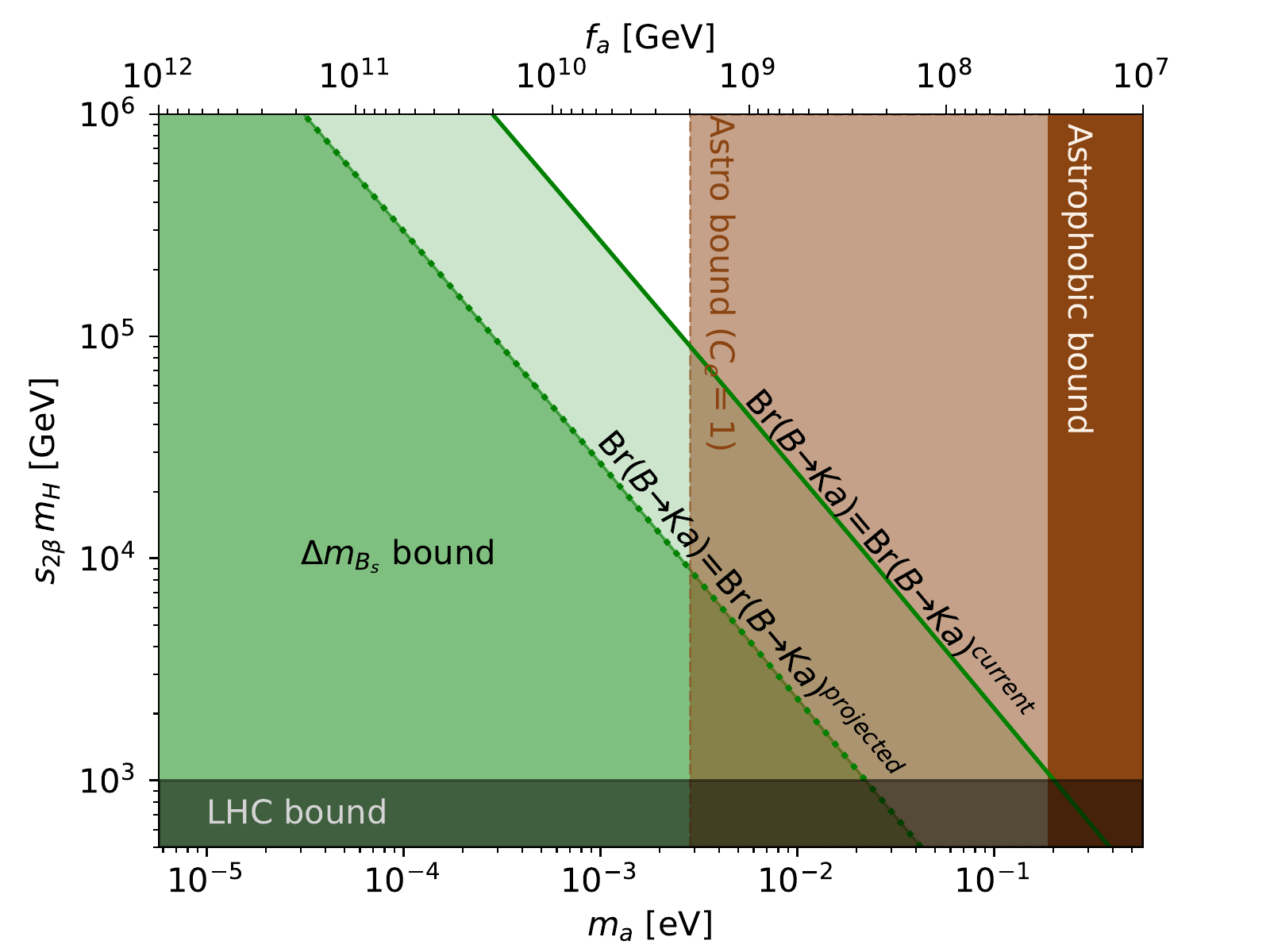}}
\end{minipage}
\hfill
\caption{Allowed parameters space after fixing the branching ratio of the axion to the current (solid), projected sensitivity (dotted) and imposing the meson mixing bound on each channel. Bounds on collider searches of 2HDM are shown in black, and astrophysical bounds for different axion scenarios are coloured in brown.}
\label{fig:UVIRplots}
\end{figure}

If an axion is found in one of these decay channels, the branching ratio is set. If we fix the branching ratio, then meson-mixing imposes a lower bound in the $f_a - m_H $ plane. This is shown in Fig.~\ref{fig:UVIRplots}, where the light coloured region is the parameter space between the current and projected sensitivity of the bounds. The dark coloured region is the bound imposed by meson mixing, the black horizontal band is the collider bound on the 2HDM and in brown we show the astrophysical bounds for a model with order one coupling to electrons, and the most astrophobic model \cite{DiLuzio:2017ogq}. This plot shows the important interplay between axion and 2HDM bounds, and we see how we could extract information about the 2HDM after detecting an axion: if the axion is found in a $s-d$ transition, then there is plenty parameter space to cover with future LHC searches. However, if an axion is found in a $b-s$ transition, then either the model is up to some extent astrophobic or the LHC would not be able to observe heavy modes of the 2HDM.

This is one particular example in which the general connection of Eq.~\ref{eq:FVconnection} can be used. Other examples, including how to relate lepton flavour-violating decays of the SM-like Higgs $h\to l_i^+ \, l_j^-$ and lepton decays $l_i\to l_j \, a$ are found in Ref.~\cite{DiLuzio:2023ndz}. We remark that these relations are most useful when there are signs of new physics, since they allow to fix at least one of the observables. 

\section{Conclusions}
In this talk we presented the results of reference \cite{DiLuzio:2023ndz}. Here, a new way of rewriting the Yukawa matrices of a non-universal DFSZ is found, allowing to link the flavour violating couplings of the heavy radial modes in terms of the axion couplings. This minimal flavour structure allows to connect flavour violating observables between the two sectors, the axion and the PQ-2HDM, that is a UV-IR connection. We have shown in an example, that information about the heavy modes can be obtained if a flavour violating axion is found. Specifically, if an axion is found in the $K\to \pi \,a$ channel then the parameter space still allows for a detection in the future runs of LHC, while if this discovery happens in other channels, e.g.~$B\to K a$, the parameter of the LHC would be already ruled out by astrophysics unless the model is astrophobic.

\section*{Acknowledgments}
This work received funding from the European Union's Horizon 2020 research and innovation programme under the Marie Sk\l{}odowska-Curie grant agreement N$^{\circ}$ 860881-HIDDeN, grant agreement N$^{\circ}$ 101086085–ASYMMETRY and by the INFN Iniziative Specifica APINE. The work of LDL is also supported by the project ``CPV-Axion'' under the Supporting TAlent in ReSearch@University of Padova (STARS@UNIPD).

\end{document}